# Observations of on-demand quantum correlation using Poisson-distributed photon pairs

Sangbae Kim and Byoung S. Ham*

Center for Photon Information Processing, School of Electrical Engineering and Computer Science, Gwangju Institute of Science and Technology
123 Chumdangwagi-ro, Buk-gu, Gwangju 61005, South Korea

*bham@gist.ac.kr

**Abstract:**

Complementarity or wave-particle duality has been the basis of quantum mechanics over the last century. Since the Hanbury Brown and Twiss experiments in 1956, the particle nature of single photons has been intensively studied for various quantum phenomena such as anticorrelation and Bell inequality violation. Regarding the fundamental question on quantumness or nonclassicality, however, no clear answer exists for what quantum entanglement should be and how to generate it. Here, we experimentally demonstrate the secrete of quantumness using the wave nature of single photons.

**Introduction**

Quantum features such as Bell inequality violation[1-6] and anticorrelation[7-10] between two-mode quantum particles or engineered individual photon pairs have been intensively studied over the last few decades to understand the quantum nature of entanglement beyond the classical limits of individuality and separability[11]. The physical understanding of quantum entanglement based on the particle nature of photons governed by the energy-time relation via the uncertainty principle has been applied for various quantum information technologies. Although the complementarity theory of quantum mechanics is normally limited to the microscopic world of conjugate entities satisfying the uncertainty principle[11,12], the Schrodinger cat[13-15] as a macroscopic quantum feature should also be confined by the same physics. Here, we report the first observation of on-demand quantum correlation using Poisson-distributed photons obtained from an attenuated laser. Since the wave nature of such Poisson-distributed photons relies on the intrinsic properties of the coherent light source, the degree of quantum correlation among photon pairs can be deterministically manipulated not only for microscopic entities, but also for a macroscopic Schrodinger cat. This paper opens the door to on-demand quantum entanglement generation from any light source.

Since the seminal paper of EPR[16] in 1935, entanglement representing the maximum correlation between two individual photons, atoms, or ions has been demonstrated via the second order intensity correlation $g^{(2)}(\tau)$, where $\tau$ is the time delay between two particles for coincidence detection in an interference scheme[7-11]. Bell inequality violation is for a nonlocal quantum feature in a noninterference scheme[1-6], while anticorrelation, the so-called HOM dip, is for the same quantum feature in an interference scheme[7-10]. If a single photon in a Fock state is considered[17], a nondefinite phase of the Fock state must be given due to the energy-time relation of the uncertainty principle. Thus, all quantum features denoted by quantum operators in a space-time domain have no definite phase relation at all. As a result, the wave nature of a single photon has been mostly forgotten in the areas of quantum information. In that sense, the terminology of coherence has been used differently depending on the context[18] ever since the temporal correlation between two individual photons was demonstrated by Hanbury Brown and Twiss[19].

Recently, a completely different approach has been pursued for anticorrelation to unveil the quantum nature of photon bunching on a beam splitter (BS)[20]. Such a pure coherence optics-based quantum theory has been successful to describe this fundamental quantum feature in terms of the wave nature of photons. As already well known, experimental results of Young's double slits or a Mach-Zehnder interferometer (MZI) show the same interference fringe whether the inputs are single photons or coherence light[21]. For single photons, the fringe formation is explained as self-interference of a single photon with itself[21]. The wave-particle duality has also been discussed for the interference scheme, where the definition of particle-like or wave-like photons is based on coherence[22-24]. According to the wave nature, however, a single photon can also be formed by superposing many coherent waves. In this coherent picture, self-interference of a single photon in Young's double-slit experiments or a Mach-Zehnder interferometer (MZI) is straightforward using the superposition principle.

Based on the wave nature of photons, further analysis of the quantum features of de Broglie waves has also been successfully introduced[25] and experimentally demonstrated[26] in the name of coherence de Broglie waves (CBW), which correspond to particle nature-based photonic de Broglie waves (PBW)[27-29]. Here, the order N in PBW is related with a hyper-entangled state whose success rate exponentially decreases as N increases[14]. Unlike probabilistic PBW, however, CBW is deterministic due to coherence, where hyper-entanglement in CBW is linearly scalable and controllable with MZI geometry. Physical understanding of CBW has been also explained using tensor products between the two-mode MZI systems[30].

**Experimental setup**

Figure 1 shows a schematic of the deterministic control of two-photon correlation as proposed in refs. 20 and 31, where individual photons are obtained from an attenuated He/Ne laser (JDS Uniphase, Novette 1508-0), whose wavelength is 632.8 nm with a 1 GHz bandwidth via Doppler broadening. With proper attenuation using neural density (ND) filters, a Poisson distributed nonclassical photon stream is obtained, where the mean photon number determines the details of the photon characteristics (see section A of the Supplementary Information). Here, the Poisson distributed photons also have the same bandwidth as the original coherent laser light regardless of the mean photon numbers. It should be noted that each photon's bandwidth is, however, homogeneous at a ~kHz linewidth determined by the decay rate of Ne atoms. For demonstration of the phase-dependent anticorrelation[20,31], the relative phase between two interacting single photons on the second BS (BS2) of MZI in Fig. 1 is controlled by PZT (Thorlabs, KCI-PZ). Regarding PZT control, however, only one axis is scanned out of three control axes to induce intentional decoherence to exhibit the classical bound as a reference, resulting in coherence-length modification of photons (see Methods and Section B of the Supplementary Information). This decoherence technique gives us an intuitive understanding for conventional measurement systems with wide-bandwidth photons as well as practical advantages in determining the two-photon correlation limit.

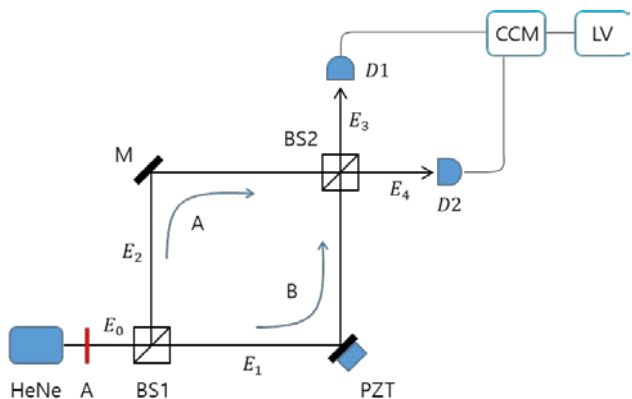

**Fig 1| A schematic of Poisson-distributed anticorrelation measurement.** A: attenuator, BS: nonpolarizing beam splitter, M: mirror. D1/D2 stands for single photon counting module SPCM; CCU: coincidence counting module; LV: Labview.

Unlike conventional anticorrelation experiments, where each entangled photon pair has a random initial phase due to the spontaneous emission process[7-10], all incident photons have the same phase, resulting in collective behavior due to the long coherence time[20,31]. Each MZI output signal ($E_3$; $E_4$) is detected by a single photon counting module (SPCM; Excelitas AQRH-SPCM-15: $D_1$; $D_2$), whose corresponding electrical signal is set at 4V with a 10 ns pulse duration (see Fig. 2). Both electrically converted single photon signals from SPCMs are fed into a coincidence counting module (CCM; Altera DE2) for both single photon counting and coincidence detection counting measurements. The resolving time-window of the SPCM is 350 ps at $\lambda = 825$ nm with a dead time of 22 ns. The dark count rate is measured as 27/s. For the results of photon counting statistics, CCM accumulates signals for $T = 1$ s.

**Results**

*Preparation*

For the Poisson distributed photon statistics, we analyze the attenuated photon statistics from the He/Ne laser using both CCM and a high-speed digital oscilloscope. For these measurements, the photon statistics is pursued for a noninterference regime by removing BS2 in Fig. 1. First, we perform photon counting measurements, analyze the obtained data, and compare them with the Poisson statistics for dozens of different OD values from the ND filter combinations (see Methods and Section C of the Supplementary Information). Second, the same measurements are performed using a high-speed (500 MHz; 400ps resolving time) digital oscilloscope (Yokogawa DL9040) to compare the visualized data with those by CCM (see Section D of the Supplementary Information). With these two different measurement techniques, we confirm that the Poisson statistics works well for the attenuated He/Ne laser and the present coincidence detection measurements. For the present coincidence detection measurements, we choose one case of ODs satisfying nonclassical photon features, whose counted numbers for single photons and bunched photons are $5.4 \times 10^5$ and 820, respectively. In this case, the mean photon number is calculated to be $\langle n \rangle = 0.012$ (see Figs. 2 and 3).

Figure 2 shows the captured data from the digital oscilloscope. The top two rows in Fig 2a show photon streams detected from both SPCMs of $D_2$ and $D_1$, respectively. Figure 2b is an expanded version of Fig. 2a for both arbitrary single and bunched photon detections. From the attenuated He/Ne laser, anti-bunched photons are randomly split into either path A or path B, resulting in no coincidence at the zero path length position. Only bunched photons are split into both paths A and B, and detected simultaneously by SPCMs. The detection rates for single and bunched photons are measured from the captured data in Fig. 2a. For this, we capture the data from both SPCMs for 1 ms, where 1 ms is the longest time to resolve single photons determined by the resolving time (400 ps each channel) of the digital oscilloscope, resulting in $2.5 \times 10^6$ resolving points in each channel. The two bottom rows in Fig. 2a are for comparison purposes to verify the present counting method through an oscilloscope, where the bottom rows come directly from a two-channel function generator (Tektronix AFG3102) with a 10 ns pulse duration and 1 MHz repetition rate.

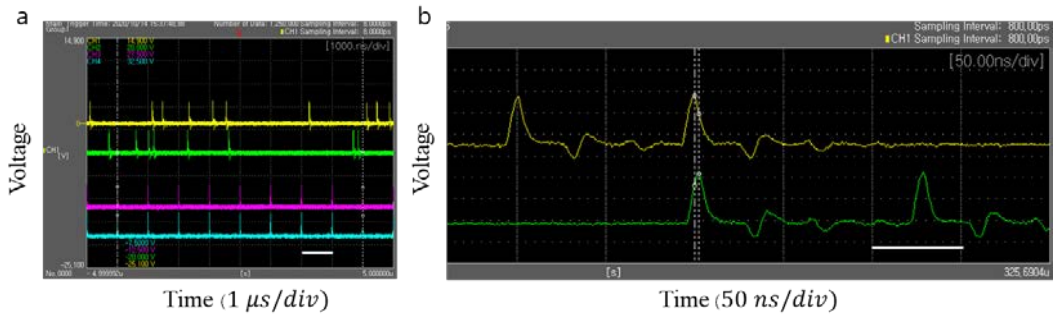

**Fig. 2| Poisson distributed photon characteristics on an oscilloscope. a,** top two rows: from each the path A and B in Fig. 1. **b,** bottom two rows: from function generator with 10 ns pulse duration at 1 MHz rate.

The measured average single photon numbers by the digital oscilloscope for the MZI paths A and B are $274 \pm 15.8$ and $265 \pm 17.6$ for T=1 ms accumulation time, respectively. Considering the dead time of 22 ns of the SPCMs, the mean photon number is calculated to be $\langle n \rangle = 0.012$: $(539)/\left(\frac{10^{-3}}{22 \times 10^{-9}}\right)$. For 16 different measurements, the standard deviation for $\langle n \rangle$ is $\sigma_{\langle n \rangle} = 17.2$. The measured counting number of the doubly bunched photons for 1 ms is $\leq 1$, satisfying the theoretical ratio of 0.005 to the single photons for the mean photon number of $\langle n \rangle = 0.012$ (Section D of the Supplementary Information). These statistical data match well with those of CCM-measured photon statistics (Section C of the Supplementary Information). Thus, we conclude that our experimental methods in Fig. 1 are valid for quantum measurements of anticorrelation. More importantly, confirmation of the bunched photons is the bedrock of the present on-demand control of quantumness via coincidence detection measurements, where a specific phase relation among the photons plays a key role.

*Coincidence measurements*

Figure 3 shows the results of Fig. 1, where the PZT is continuously scanned for 316 seconds (0~100V) manually across the zero path-length difference, satisfying a ~1 s accumulation time for each data point. Figure 3a shows the experimental data from Labview (LV) fed from CCM for single photon counting measurements as a function of the PZT scan time across the zero path length at the center position. The black and red dots are for the SPCM measurement data from path A and B, respectively. As discussed already[31], the modulation period of each fringe pattern is due mostly to single photons due to extremely low rate for bunched photons, and thus confirms the self-interference. The fringe period represents the wavelength of the He/Ne laser at 632.8 nm, where the slightly different period is due to the manual scanning. Considering the dead time of SPCMs, the mean photon number is calculated to be $\langle n \rangle = 0.012: \frac{5.4 \times 10^5}{4.5 \times 10^7}$ (see section C of the Supplementary Information). As a reference, the same $g^{(1)}$ correlation is separately measured for an unattenuated He/Ne laser in Fig. 3b, where the detector is replaced by typical Si photodiodes (Thorlabs, APD110A). Figures 3a and 3b demonstrate no discrepancy between single photons and coherence light in MZI physics. This also shows that the present antibunched photons relied on the wave nature of the photons.

To include the incoherence feature in coincidence detection measurements as a reference, we intentionally modify the MZI system by applying a simple asymmetric PZT scanning method for path B, resulting in a cross section walk-off on BS2 (see Methods and Section B of the Supplementary Information). This cross section walk-off causes PZT scan length-proportional phase shifts, resulting in wide bandwidth-photon-like characteristics. Both ends of the self-interference fringes in Fig. 3a demonstrate this walk-off caused incoherence feature compared with the center fringe of the original coherence-based one. Compared with the original coherence length of 30 cm, the fringe envelopes observed in Figs. 3a and 3b indicate a shortened coherence-length effect at ~2 μm. This effect is not due to actual decoherence in each photon, but instead the cross section walk-off-based artifact on BS2. Thus, the attenuated He/Ne laser is proved to fit Poisson statistics with spectral modification by the asymmetric PZT scanning method. As a result, both far ends of the interference in Fig. 3a can be treated as a classical lower bound of $g^{(2)} = 0.5$[20,24].

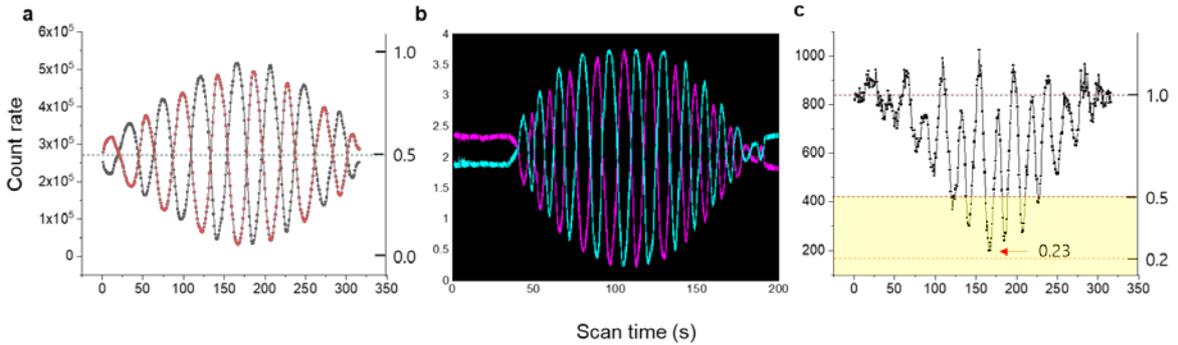

**Fig. 3| Experimental results for Fig. 1. a,** Single photon counting rates: $N_A$ (black dots); $N_B$ (red dots). **b,** Amplitude correlation, $g^{(1)}$ with coherent light. **c,** Coincidence counting rate. Right axis: normalization. A: PZT related path length change (arbitrary). The path length corresponds to the manual scanning time of the PZT controller.

The maximum fringe visibilities at the center for the zero path-length difference are calculated to be 88.2% and 87.7% for the black and red curves, respectively. The non-perfect visibility in Fig. 3a is due to non-perfect overlap-caused errors on BS2 in MZI. This non-perfect overlap is caused by nonperfect alignment, atmospheric turbulence, and/or nonperfect reflection/transmission ratio of BSs. The measured maximum visibilities at line center are beyond the classical limit of 70.7% according to the Bell inequality[1-6]. The measured fringe visibility by coherent light in Fig. 3b is 90.0%, which is nearly the same as in Fig. 3a considering the standard deviation. Unlike most quantum measurements, the group of photons due to the PZT control behaves collectively because of the wave nature if there is no asymmetric scanning. This is because there is no difference in the fringe patterns between single photons in Fig. 3a and coherence light in Fig. 3b. The right vertical axis of Fig. 3a indicates a normalized counting rate, where the midpoint of 0.5 represents the classical lower bound of incoherence (individuality or particle nature) as a reference.

Figure 3c shows a raw coincidence counting rate for Fig. 3a, where the coincidence counting rate, corresponding to $g^{(2)}$, is doubly modulated compared with that in Fig. 3a for $g^{(1)}$, as expected[20,31]. The coincidence detection measurements in Fig. 3c is of course not for the single photons but for the bunched photons of $E_0$ in Fig. 1, because coincidently split photon pairs in MZI are required by definition. The ratio of bunched photons to single photos is measured as $\eta_{21} = 0.0015$: $\eta_{21} = \frac{820}{270,000 \times 2}$ (see section C of the Supplementary Information). The observed modulation frequency for the coincidence detection rate is new and unprecedented. By the definition of the AND logic gate in CCM, each crossing point between the red and black curves in Fig. 3a gives maxima in Fig. 3c. The alternative discrepancy in maxima may be due to the errors of the asymmetric PZT scan method. On the contrary, each anti-crossing point between the red and black curves in Fig. 3a gives minima in Fig. 3c. Because the coincidence detection is strongly related with the intensity correlation $g^{(2)}$, all values of coincidence in Fig. 3c must lie between zero as a minimum and one as a maximum for coherent lights (see the right vertical axis). Here, three or more bunched photons are negligible according to the Poisson statistics (see Section A of the Supplementary Information). Any value below the mid-value of the coincidence rate in Fig. 3c indicates a quantum feature due to two-photon interactions and proves entangled photon pairs (see the shaded area), where the lowest value of the intensity correlation is $g^{(2)}(0) = 0.24$ (discussed in Fig. 4): $\frac{200}{820}$. Such a nonclassical feature in Fig. 3c is phase dependent, as expected in refs. 20 and 31, resulting in periodic fringe (see Discussion). Thus, the on-demand nonclassical feature of anticorrelation is achieved when a specific relative phase is met between an impinging photon pair at $\pm\pi/2$. Such a $\pm\pi/2$ phase shift is the origin of entangled photon pairs[20,31]. Thus, the conventional probabilistic quantum feature in $g^{(2)}$ correlation is now replaced by a deterministic one under the wave nature of photons.

## *Discussion for $g^{(2)}(\tau)$ in Fig. 3c*

According to the definition of intensity correlation, $g^{(2)}(\tau) = \frac{\langle I_1 I_2(\tau) \rangle}{\langle I_1 \rangle \langle I_2(\tau) \rangle} = \frac{\langle N_1 N_2(\tau) \rangle}{\langle N_1 \rangle \langle N_2(\tau) \rangle} \frac{T}{\delta t}$, the denominator is composed of mean intensity product of each input channel (A and B). In photon statistics with a particle nature, the bunched photon output of BS2 is directed into either D1 or D2 with equal chance due to the undecided sign of $\pm\pi/2$ in the SPDC generated entangled photon pairs[7-10]. Thus, $g^{(2)}(\tau)$ is proportional to the coincidence detection rate.

On the contrary, in the present experiments based on the wave nature of photons, the output port of the bunched photons on BS2 is definitely determined if the input phase relation is fixed. As a result, the denominator cannot be uniform for the relative phase and varies with the coincidence detection rate. For example, for a zero phase difference between paths A and B, only detector D2 clicks regardless of the input photon's frequency and phase. As mentioned above, the bunched input photons ($E_0$) are split into A and B equally, and the relative phase of $\pi/2$ is achieved by BS1[32]. Unlike entangled photon pairs via SPDC processes, the sign of $\pi/2$ is predetermined by the BS. Thus, the MZI output direction by BS2 for the split photons is also determined. The resulting minima of the coincidence detection rate are repeated at every half wavelength or $\pi$ phase shift, as shown in Fig. 3c.

The phase dependent $g^{(1)}$ in the output ports of MZI is opposite each other, and thus, the product $I_1 I_2(\tau)$ should be exactly the same as the coincidence detection rate, resulting in $g^{(2)}(\tau) = 1$ regardless of $\tau$ or the phase difference if $g^{(1)} = 0$ is excluded. To circumvent this, an opposite case of Fig. 3a can be considered by adding an additional $\pi-$phase to PZT. This additional case results in swapping between the red and black curves in Fig. 3a, but makes no difference in the coincidence detection measurements in Fig. 3c. Thus, $g^{(2)}(\tau)$ can be calculated by averaging these two cases. Obviously, this averaging effect makes the present system conceptually equal to conventional particle nature-based intensity correlation. Here, the predetermined MZI output does not violate the uncertainty principle as the fixed sum frequency in SPDC processes.

Figure 4 provides numerical simulations for Fig. 3a and 3c with perfect overlap corresponding to the asymmetric PZT scan method in Fig. 1. The intentional decoherence effect observed in Fig. 3a representing the classical lower bound of $g^{(2)} = 0.5$ is quickly simulated by multiplying a Gaussian envelope G to the perfect coherence fringe, as shown in Fig. 4a, where the Gaussian function implies a decoherence-caused spectral bandwidth of photons. The related coincidence counting rate is also obtained directly from Fig. 4a by multiplying the red and blue curves as shown in Fig. 4b, which is similar to Fig. 3c. With the technique circumventing the zero denominator in $g^{(2)}(\tau)$ as mentioned above, the intensity correlation $g^{(2)}(\tau)$ can be obtained from Fig. 4b, where the decoherence-induced classical feature of $g^{(2)}(\tau) = 0.5$ is simply added by multiplying the same Gaussian function (see Fig. 4c). In Fig. 4c, it is clear that the phase control of MZI generates all classical and quantum phenomena depending on $\varphi$. In an ideal condition of $l_c \gg \lambda$ without the decoherence effect, the intensity correlation $g^{(2)}(\tau)$ varies from 0 to 1 as discussed in refs. 20 and 31. In that sense, the minimum value of the coincidence detection rate in Fig. 3c is treated as $g^{(2)}(0) = 0.24$ ($\frac{200}{820}$), which demonstrates an entangled photon pair.

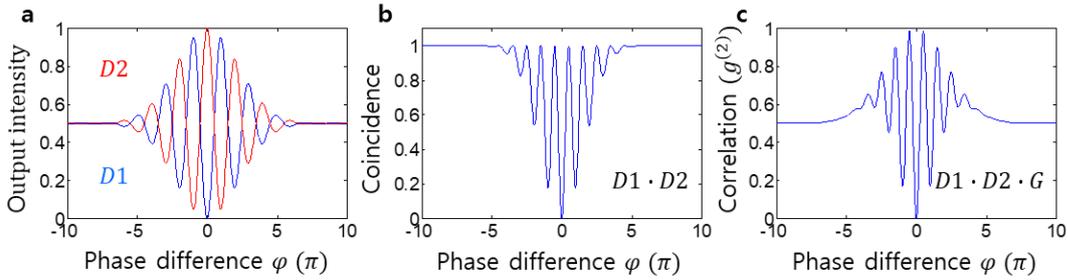

**Fig. 4| Numerical simulation for intensity correlation $g^{(2)}(\tau \sim 0)$ for Fig. 3c. a,** Normalized Output intensity applied by a Gaussian function. **b,** Normalized coincidence detection **c,** Intensity correlation based on a. G: Gaussian envelope.

**Discussion**

Heisenberg's uncertainty principle is the crux of quantum mechanics, where wave-particle duality is conceptually another expression of this principle. In quantum measurements using the particle nature of photons represented by two orthonormal bases such as polarization (vertical and horizontal) and phase (0 and $\pi$), the probabilistic treatment of photons with respect to the corresponding conjugate variable is necessary. In most quantum measurements, however, a photon's coherence property is pre-determined at photon generation in the medium. In most SPDC-generated entangled photons, the bandwidth is ~5THz, whose corresponding coherence time (length) is $t_c = 200$ fs ($l_c = 60$ $\mu m$). Thus, photon coherence is still enough for the PZT scan range. As discussed in ref. 31, the disappearance of $g^{(1)}$ in the so-called HOM dip is due to the random phase-caused coherence washout among entangled photon pairs via the spontaneous emission process. On the contrary, in the present attenuated laser system, all photons are coherent with the same initial phase. Thus, the $g^{(1)}$ term survives in the coincidence detection measurements. The same physics of coherence (the wave nature of photons) is shared in both systems, but results in opposite outputs as either nondetermnistic control for the SPDC case or deterministic control for the coherent case. Most of all, the photon number does not affect the results for the coherent case.

**Conclusion**

We experimentally demonstrated the novel feature of on-demand two-photon correlation corresponding to $g^{(2)}(0) = 0.24$, whose manipulation is not probabilistic but deterministic. For experimental demonstrations of this deterministic anticorrelation, we used Poisson distributed photons obtained from an attenuated He/Ne laser. With several OD values of the ND filters, we measured the photon statistics of the attenuated laser, and bunched photons were used for coincidence detection measurements. For this, a simple MZI scheme was adapted as a coherence quantum device, where the phase difference between two photons located in two different paths was precisely controlled by PZT. To induce an artificial incoherence effect between the coherent photon pairs, a single-axis control method for PZT scanning was adapted, otherwise a long coherence time was kept for the coincidence detection. For coincidence detection between two output photons of MZI, we confirmed that only bunched photons contributed to the coincidence detection. Unlike conventional coincidence detection measurements with entangled photon pairs or spectrally engineered photon pairs, the observed two-photon correlation was phase controllable and covered all degrees of freedom between quantum and classical regimes. Thus, the observed coincidence detection fringes corresponding to $g^{(2)}(\tau) < 0.5$ demonstrated that the quantum feature is phase-controllable as proposed in refs. 20 and 31. Moreover, collective photon behavior can be implemented for macroscopic quantum features of such as a Schrodinger cat or N00N state generations.

**Methods**

To set a zero-path length difference between the two paths of MZI in Fig. 1, a standard interference test is performed using an unattenuated He/Ne laser. After setting the PZT scan position at the zero path length, BS2 is removed to avoid any interference effects. Then, the CCM is replaced by a fast (500 MHz; 400 ps resolving time for two channels) four-channel digital oscilloscope (YOKOGAWA DL9040) to visualize each photon's characteristics. The size of the MZI is ~$10 \times 10 cm^2$, enclosed by a cartoon box to minimize air fluctuation-caused phase noise. Because the bandwidth-based coherence length of the He/Ne laser is 30 cm, the PZT-caused MZI interference fringe is well defined within the coherence length, whose PZT scan range (resolution) is ~$\pm$ 4 μm (0.5 μrad). The photon stream from the attenuated He/Ne laser is sent to the MZI setup in Fig. 1 to split the photon stream into two paths of A and B via BS1. Because each group of photons in A or B behave collectively due to the wave nature given by the He/Ne laser, the present scheme can also be applied for a macroscopic regime[25-31].

1. Dark count measurement for SPCM

For a dark room condition in a cartoon box, both single photon counting modules (SPCM-AQRH-15) result in the dark count rate at $27 \pm 5$ (count/s) without input photons, which is satisfied with the manufacture specification at 50 (count/s). The same rate is also confirmed with blockage of the photodetector.

2. Attenuated He/Ne laser

The attenuated light source is He/Ne laser (Thorlabs; PM100D). In free running for two hours, the output intensity is stabilized at $136 \pm 0.16$ μW. The ND filters are applied to this stabilized output for the experiments in Figs. 2 and 3. The optical density varies from 8.74 to 14.9, whose corresponding measured powers are between 0.2 aW to 0.29 pW.

3. Counting method for single and bunched photons on a coincidence counting module

The SPCM-generated electrical pulses are sent to the coincidence counting module (DE2 FPGA, Altera). The electrical pulse duration from SPCM is ~10 ns, and coincidence detection is counted for about half overlapping between two pulses in DE2. Both single and coincidence counting numbers are measured in parallel for 100 ms at each step and the measured data is transferred to PC through RS232 cable for a coincidence_rs232(4_5).vi application program in real time. By setting the counting period at one second, the number of photons is calculated at every second by the vi program and recorded by the coincidence_recorder_rs232(4_5).vi program in real time. For the coincidence measurements in Fig. 3c, the number of bunched photons generated by the attenuated He/Ne laser was measured for all different ODs (see Section D of the Supplementary Information).

4. Counting method for single and bunched photons on a fast digital oscilloscope

The ND filters are set for 270,000 (count/s) in each path of MZI in Fig. 1, where the single photon counting number is confirmed by the coincidence counting module, DE2 FPGA. Then, the SPCMs are directly connected to the four-channel digital oscilloscope (YOKOGAWA; DL9040). Channel 1 of the oscilloscope is for the path B, while Channel 2 is for the path A. The time scale of the oscilloscope is set for the maximum resolution of 100μs per division, where the full screen is for 1 ms, and the number of pulses are 270 per channel. To measure bunched (coincidence) photons between two channels, a homemade MATLAB program is used. For 16 data samples, the average number of pulses for channel 1 and channel 2 are 274.9 and 265.2, respectively, where standard deviations are 15.8 and 17.6. We have also confirmed the program results with visual counting numbers on the oscilloscope.

  5. Asymmetric PZT scanning

This coherence-length modification is directly resulted from the linearly increased phase shift on a cross section of the overlapped photons for a fixed PZT scan position, otherwise a long and uniform coherence length. The PZT is a 3-channel open-loop piezo controllers (Thorlabs; MDT693B) attached to a mirror as shown in Fig. 1 to adjust variation of overlap cross section on BS2. For the PZT scan, the voltage range varies from 0 to 100V whose voltage resolution is 1.5 mV. We manually scan the PZT controller. For the symmetric walk-off on the cross section, the PZT controlled beam overlap is to be perfect in the middle of the voltage range.

**Data availability**

Data for figures are available upon requests.

**Acknowledgments** This work was supported by GIST via GRI 2020.

**Author contributions** B.S.H. conceived the research on both theory and experiments, did numerical simulations, data analysis, and wrote the manuscript; S.K performed the experiments and analyzed the data.

**Correspondence and request of materials** should be addressed to BSH (email: bham@gist.ac.kr).

**Competing interests** The author declares no competing (both financial and non-financial) interests.

**Supplementary information** is available in the online version of the paper.